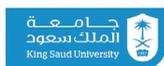

Contents lists available at ScienceDirect

# Journal of King Saud University – Computer and Information Sciences

journal homepage: www.sciencedirect.com

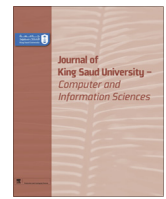

# STORE: Security Threat Oriented Requirements Engineering Methodology

Md Tarique Jamal Ansari [a,*], Dhirendra Pandey [a], Mamdouh Alenezi [b]

[a] Department of Information Technology, BBA University, India
[b] College of Computer & Information Sciences, Prince Sultan University, Saudi Arabia

## ABSTRACT

As we are continuously depending on information technology applications by adopting electronic channels and software applications for our business, online transaction and communication, software security is increasingly becoming a necessity and more advanced concern. Both the functional and non-functional requirements are important and provide the necessary needs at the early phases of the software development process, specifically in the requirement phase. The aim of this research is to identify security threats early in the software development process to help the requirement engineer elicit appropriate security requirements in a more systematic manner throughout the requirement engineering process to ensure a secure and quality software development. This article proposes the STORE methodology for security requirement elicitation based on security threats analysis, which includes the identification of four points: PoA, PoB, PoC and PoD for effective security attack analysis. Further, the proposed STORE methodology is also validated by a case study of an ERP System. We also compare our STORE methodology with two existing techniques, namely, SQUARE and MOSRE. We have shown that more effective and efficient security requirements can be elicited by the STORE methodology and that it helps the security requirement engineer to elicit security requirements in a more organized manner.
© 2018 The Authors. Production and hosting by Elsevier B.V. on behalf of King Saud University. This is an open access article under the CC BY-NC-ND license (http://creativecommons.org/licenses/by-nc-nd/4.0/).

## 1. Introduction

In recent years, several organizations are becoming heavily dependent on information processing systems to get more benefit quickly. This dependency has produced a need for protecting software systems from threats. Therefore, software security has become an essential issue (Zulkernine and Ahamed, 2006; Ansari and Pandey, 2018). Software security vulnerabilities and flaws are outcomes of poorly built software that can lead to easy exploitation by the hackers. Inappropriate security requirements engineering is one of the reasons for developing bad quality software products. This problem can be reduced by taking security concern into account from the early phases of the software development process.

The development process needs to shape their security properties by adding security methodology and implementing them correctly, by observing security principles, and by avoiding defects. Security requirements guide this design, implementation, and verification work (Türpe, 2017). Gartner, Inc. forecasts worldwide enterprise security spending to total $96.3 billion in 2018, an increase of 8 percent from 2017. This report shows that the digital industries are spending huge money on maintaining security consequences of regulations, changing and attracting customer mindset, awareness of emerging threats and the advancement to a digital industry planning (Gartner, 2018).

The White House cybersecurity coordinator (Rob Joyce, 2017) has outlined that receiving and maintaining the essential cyber capabilities to protect the nation creates a tension between the government's need to maintain the resources to follow attackers in the digital world through the use of digital exploits, and its obligation to share its knowledge of flaws in software and hardware with responsible parties who can ensure digital infrastructure is upgraded and made stronger in the face of growing cyber threats. Amoroso (2018) argue that the cybersecurity community, in particular, has been vocal about finding ways to improve the software, because most vulnerabilities involve exploitable weaknesses introduced through badly written code. Security requirement engineering is a vital phase in the software development process which ensures secure software development from the beginning. In the last few years, the security has become a crucial issue in the development and deployment of secure software products. This crucial issue has been considered by several security professionals and researchers for publishing research papers on integrating security considerations into the software development process. Yet, only a small amount of papers illustrates complex case studies (Massacci et al., 2005). To solve the security problem with software and web-based applications, an increasing number of literature publication, symposium tracks, and workshops in recent years call attention to the increasing attention of researchers and authors in designing and developing security requirements engineering frameworks, techniques, process, and methodologies.

\* Corresponding author.
*E-mail addresses:* tjtjansari@gmail.com (M.T.J. Ansari), malenezi@psu.edu.sa (M. Alenezi).

Security requirement engineering relates to non-functional requirements of the software system which is a very important task to achieve for a quality software product. Improving the quality of requirements is thus important. But it is a difficult objective to achieve. To understand the reason, one should first define what requirements engineering is really about (Van Lamsweerde, 2000). Security requirements of computer-based systems, mostly concern confidentiality, integrity, and availability features (Herrmann and Herrmann, 2006). Security is considered as a balance between confidentiality, integrity, and availability (Olivier, 2002). Security requirements are a type of non-functional requirements (Devanbu and Stubblebine, 2000). According to (Firesmith, 2004a, b) security requirement enhanced the quality of the software product by adding a necessary quantity of security regarding conditions related to a system and a low level of an associated quality measure that is essential to convene one or extra security policies. Engineering software security is an analysis of identifying security issues for a software product in a systematic approach. This is a very important process taken into consideration at an early stage of the software development process for the achievement of the secure software product. Security requirements are necessary and it takes place when the stakeholders found that some objects in the context of the software system are precious, they might be tangible or intangible, they have some value that the stakeholders desire to protect (Haley et al., 2006).

Several researchers have proposed security requirements engineering techniques, tools, framework, methodology, and norms for eliciting security requirements during the early stages of the software development cycle. Engineering security requirement into the early stages of the development process is profitable, secure and also provides quality software product. Security requirement engineering should be systematic, repeatable and capable of eliciting complete, reliable, clear, simple and easy to analyzable by the other members of the software development process (Kotonya and Sommerville, 1998). Security requirements are often developed independently of other requirements engineering activities. Generally, security requirement engineering carried out separately of other functional requirements engineering activities. Therefore, many vital security requirements are repeatedly ignored, and the requirement engineer only focused on functional requirements by ignoring important security aspects (Mead, 2008). Several security requirement engineering frameworks (Yu,1997; Moffett and Nuseibeh, 2003; Viega, 2005; Lee et al., 2006; Tsoumas and Gritzalis, 2006; Hussein and Zulkernine, 2007; Haley et al., 2008; Salini and Kanmani, 2013; Riaz et al., 2016; Ansari and Pandey, 2017; Rehman and Gruhn, 2018), techniques (Jürjens, 2002; Popp et al., 2003; Peeters, 2005; Firesmith, 2005; Sindre and Opdahl, 2005; Basin et al., 2006; Paja et al., 2015), processes (Toval et al., 2002; Lee et al., 2003; Zuccato, 2004; Myagmar et al., 2005; Mellado et al., 2007; Van Lamsweerde, 2007; Shin and Gomaa, 2007; Hassan et al., 2010; Mufti et al., 2018) and methodologies (Bresciani et al., 2004; Jennex, 2005; Mead and Stehney, 2005; El-Hadary and El-Kassas, 2014) have been proposed by different authors.

Normally, Requirements engineers are well skilled in functional requirements, but not in software security. Very few requirement engineers that have been skilled have only been known the basic security architectural knowledge like password security, encryption, and decryption. They don't have deep knowledge of authentic security requirements engineering. The software system which is to be developed may have several stakeholders. Only some stakeholders participated in the development of software products, but all stakeholders are associated with the software system. Some stakeholders are capable to help in the identification process of the asset. They suggest assets of the system from their point of view. These assets may have vulnerabilities. The threat exploits these vulnerabilities to get access to the system. The risk associated with each identified threat is not the same. The attacker has always targeted the vulnerability with high risk. The requirement engineer categorizes and prioritizes the identified threats. After that, threats are mitigated by using some security mechanism which fulfills the security requirements of the system.

The elicitation of effective and efficient security requirements is an important and challenging task. It is an important task because there are many problems associated with the consideration of security issues during the software development phases that must be overcome. Security requirement engineering is challenging because there are requirements that such a security-oriented approach must satisfy. Generally, software security is not considered by the developers during the early phases of the software development life cycle (Ansari and Pandey, 2017).

Most security Requirements engineering approach normally does not comprise all significant stakeholders and does not use the well-organized techniques for stakeholder identification and prioritization (Faily, 2015). Normally the security requirements specification is incomplete, confusing, conflicting, not cohesive, disorganized, infeasible, obsolete, unable to be validated, and not usable by their anticipated persons (Mead, 2008). Several authors have been proposed methodology for analyzing security requirements engineering. All effective and efficient security requirements should be well organized in a systematic manner otherwise the software system cannot be evaluated for accomplishment. Therefore, there is a need to develop a framework which is capable of eliciting more effective and efficient security requirements by considering all the issues which are neglected by the previous approaches.

The main aim of this paper is to design and develop a security threat oriented security requirements engineering methodology that is capable of eliciting security requirement which is effective, efficient, complete, clear, consistent, organized, feasible, up to date and easy to be validated. This methodology is especially suitable for any type of software product that requires security from the beginning of the development process. This paper presents a security threat oriented requirement engineering (STORE) methodology which is a ten step systematic process. Here we describe every step and also the participants involved with every step. Our main aim in the proposed methodology is to get as many effective and efficient security requirements as possible for the secure and quality software product development.

The organization of this paper is as follows. Section 2 "Related work" discusses several security requirements engineering approaches. Section 3 "Proposed Approach" presents our proposed methodology for security requirements engineering. Section 4 "Case study" presents the implementation of our proposed methodology on a case study of ERP web-based application software. Section 5 "Results and discussion" compares results of applying our proposed methodology with two related approaches. Section 6 "Conclusion" summarizes our work.

## 2. Related work

Several requirements engineering approaches have been developed in order to develop a secure software product. Each approach represents unique features that make it syntactically and semantically different from the other one. (Toval et al., 2002) presented a practical approach for managing the security of the information systems from the requirement engineering process. This approach is a particularization of a general-purpose process for requirements reuse called SIREN, which shortens the development process since the analyst starts from a reusable set of requirements. (Jürjens, 2002; Popp et al., 2003) presented the extension of UMLsec of UML to add security-related information within the UML diagram in a system specification. This security requirements model has multilevel security and compulsory access control. (Basin et al., 2003) have proposed a model-driven security approach. For that, they specialized model-driven architecture paradigm into model-driven security. After that, they designed an application for developing software systems through process models, in which they integrate the process design language UML with a security modeling language SecureUML for assigning access control requirements. The process models in the integration of UML and SecureUML are used to repeatedly produce security architectures for distributed software applications.

System-theoretic considerations based approach (Zuccato, 2004) was proposed for the elicitation of security requirements. This approach shows that security requirements can be described with the aid of analysis in the business environment, meeting with all the potential stakeholders and risk investigation. The Tropos methodology (Bresciani et al., 2004) is anticipated to make the analysis and design tasks easier in the software development process. Tropos approach is based on ranking and revised components of the i* framework. Tropos approach based on the concept of using requirements modeling idea of making a model of the system-to-be within its operational environment, which is incrementally sophisticated and extensive to provided that both a common interface for a variety of software development tasks and as a foundation for the documentation and development of the software product. The development phases of Tropos approach are mainly initial requirement, late requirements, architectural design, complete design, and implementation. Several authors (Giorgini et al., 2006; Mouratidis and Giorgini, 2007; Ali et al., 2009; Dalpiaz et al., 2009) have proposed the extensions of Tropos. (Giorgini et al., 2006) is one of the most significant extensions which present a formal framework for analyzing and modeling security and trust requirements.

Further (Jennex, 2005) proposed a methodology which is based on meta-notation to insert security information to access system development diagrams. This approach considered the technique of integrating security design into the software development lifecycle. Another author (Firesmith, 2003; Firesmith, 2004a,b; Firesmith, 2005) proposed a technique with some steps which define security requirements from reusable templates. He has performed a security analysis which is based on two fundamental concepts acquired from OCTAVE (Operationally Critical Threat, Asset, and Vulnerability Evaluation). He proposed security use cases as a method that should be used to identify the security requirements that the applications will effectively accomplish to secure themselves from the significant security threats in the software development process. SQUARE (Mead and Stehney, 2005) is a methodology which consists of nine step systematic security requirements engineering methodology. This methodology delivers a technique for eliciting, categorizing, and prioritizing security requirements for IT application and software product. The SQUARE methodology is mainly used for building security conceptions in the early phases of the software development lifecycle. The SQUARE methodology may also be used for documenting and evaluating the security requirements of software systems. Another author Peeters (2005) extends the agile practices to solve the security issues in an informal, open and assurance determined way. With the intention of rising the agility of requirement engineering, He sets, advancing the concept of using "abuser stories". These stories identify how the attackers may abuse the system and threaten stakeholder's assets. The abuser stories thus make the establishment of security requirements easier.

Sindre and Opdahl (2005) extend the traditional use case approach to also consider misuse cases, which signify the functionality not wanted in the system to be developed. Viega (2005) proposed a subset of CLASP, which is a set of process pieces for helping development organizations improve the security of their software. The basic idea behind the way that CLASP handles security requirements is the performance of a structured walk-through of resources, determining how they address each core security service throughout the lifetime of that resource. Another approach which is based on threat modeling for security requirement elicitation done by Myagmar et al. (2005) investigate how threat modeling can be used as foundations for the specification of security requirements and they also present three case studies of threat modeling.

Tsoumas and Gritzalis (2006) designated a security framework of an arbitrary information system. This is a security ontology-based framework which extends the DMTF Common Information Model (CIM) with ontological semantics with the intention of practicing it as a container for information system security associated info. This framework delivers security attainment and knowledge management in an effective way. Gürses (2006) introduced MSRA (Multilateral security requirements analysis) method that integrates the process of eliciting security requirements of the end-users into the requirements elicitation process of a multilaterally secure system. The aim of this approach is to identify and analyze security requirements from the multiple views of stakeholders. Basin et al. (2006) have proposed a model-driven approach for building secure software system and implemented this approach in a UML-based CASE tool.

Hussein and Zulkernine (2007) have proposed a framework for developing components with intrusion detection capabilities. In this framework, the first step elicits the requirements, in which the developers detect services and intrusions. Specifically, they capture user requirements concerning the services and functionalities provided by the components and identify the unwanted or illegal usage of components by intruders. Intrusion scenarios are elicited through the use of additional approach misuse cases of a UML profile called UMLintr. Van Lamsweerde (2007) extends KAOS to comprise the embellishment of security requirements. His proposed approach is a model based and to a certain extent relies on the use of formal methods when and where needed for RE-specific tasks, notably, goal refinement and operationalization, analysis of hazards and threats, conflict management, and synthesis of behavioral models. Mellado et al. (2007) proposed a standards-based process, named SREP (Security Requirements Engineering Process) that deals with the security requirements during the early stages of software development in a systematic and intuitive way.

Haley et al. (2008) proposed a framework for security requirements elicitation and analysis. This framework is based on creating a background for the software system, demonstrating security requirements as constraints, and elaborating satisfaction arguments for the security requirements. Hassan et al. (2010) proposed a formal analysis and design for engineering security (FADES) as the first goal-oriented software security engineering approach.

This approach provides an automated bridge between the goal-oriented semi-formal Knowledge Acquisition for automated Specifications (KAOS) framework and the B formal method. Salini and Kanmani (2013) proposed a Model oriented framework to Security Requirement Engineering (MOSRE) framework that uses a use case diagram to elicit security requirements. MOSRE has been applied to E-Health web applications. To determine the security requirements, it has the ability to identify, quantify and rank the risks of security threats and vulnerabilities. El-Hadary and El-Kassas (2014) proposed a methodology based on problem frames and abuse frames for security requirement elicitation. They used problem frames to build a security catalog and to represent security requirements, while the abuse frames are used for threat modeling. Paja et al. (2015) proposed STS approach for modeling and reasoning about security requirements. In this approach, security requirements are identified by the STS-ml requirements modeling language. The requirements models of STS-ml have a formal semantics which allows automated reasoning for detecting possible conflicts among security requirements along with conflicts between security requirements and actors' business policies. Riaz et al. (2016) developed Discovering Goals for Security (DIGS) framework, that models the key entities in information security, including assets and security goals. Ansari and Pandey (2017) proposed a framework by integrating three effective security requirements elicitation techniques, Threat modeling, Misuse case and Attack pattern for the elicitation of security requirements.

Recently, authors have discussed and proposed new approaches to security requirements engineering. Mufti et al. (2018) developed the Requirements Engineering Readiness Model (SRERM) to allow organizations to measure their security requirements engineering (SRE) readiness levels. They conducted two case studies to measure the usability of the SRERM. Rehman and Gruhn, (2018) proposed security requirements framework for CPSs that overcomes the issue of security requirements elicitation for heterogeneous CPS components.

Several research papers have been published that describes security requirements engineering in terms of methodology, techniques, process, and frameworks. Most security requirements engineering methods consider the complete CIA triad. Some of them also address other requirements than security requirements. Although considering stakeholders view in security requirement engineering is an important concern, but only some SRE approaches like MSRA, KAOS, Secure Tropos, SQUARE, STS, DIGS, and SRERM address this concern. This doesn't mean that it is impossible to consider the views of different stakeholders using other methods. However, most of the security requirements do not capture this issue in their various activities. MSRA (Gürses and Santen, 2006) is the only security requirements engineering approach that proposes steps to establish a compromise between different security concerns accepted by different stakeholders. All stakeholders incorporate during this process (Fabian et al., 2010). The concept of a counter-stakeholder in MSRA cannot be considered a threat agent, because it does not imply that the counter-stakeholder will threaten the system. SecureUML is concerned with access control only. Therefore, general threat analysis is out of the scope of this method. The proposed STORE methodology provides steps for considering security interests of all the stakeholders of the software system and also involves the threat agent identification and risk analysis for easy, effective and efficient security requirements engineering.

## 3. Proposed approach

The following sections describe the proposed STORE methodology and its several steps in detail.

### 3.1. Security Threat Oriented Requirements Engineering (STORE) methodology

The main contribution of this paper is the proposed STORE methodology which is our novel work. The STORE methodology is a ten-step sequential process which provides an effective, efficient and systematic way of eliciting and documenting security requirements for the software as well as web-based applications from the early phases of software development. The STORE methodology provides steps for considering security interests of all the stakeholders of the software system and also involves the threat agent identification and risk analysis for easy, effective and efficient security requirements engineering. In this methodology, security requirements are often discussed in the context of threats. Threat helps the security requirement engineer to calculate the risk associated with it and also represents the adversary's abilities. Stakeholder plays an important role in the STORE methodology. The following Fig. 1 shows the security requirements engineering concept model for the proposed STORE methodology where SH represents stakeholder and A represents corresponding asset.

A software system which is to be developed can have several stakeholders, only a few stakeholders are associated with the security of software products. Those stakeholders, who have security concern of the software system, have knowledge about the related assets of the systems which are to be protected from the threats. The following Fig. 2 shows an activity diagram of STORE methodology.

In STORE methodology we identify and prioritize all such stakeholders based on their importance. It is important to consider every significant stakeholder from the beginning of software development. The proposed methodology considers security threats for identifying security requirements with the help of potential stakeholders. These stakeholders help the requirement engineer in asset identification of the software product. The STORE methodology starts with identifying system goals. Each step of STORE methodology is equally important. The requirement engineer can't skip or jump to any other step because it is a systematic approach for eliciting and documenting security requirement. The STORE methodology is completely based on threats because after identify-ing potential threats the requirement engineer is capable of elicit-ing security requirements with the help of threat dictionary.

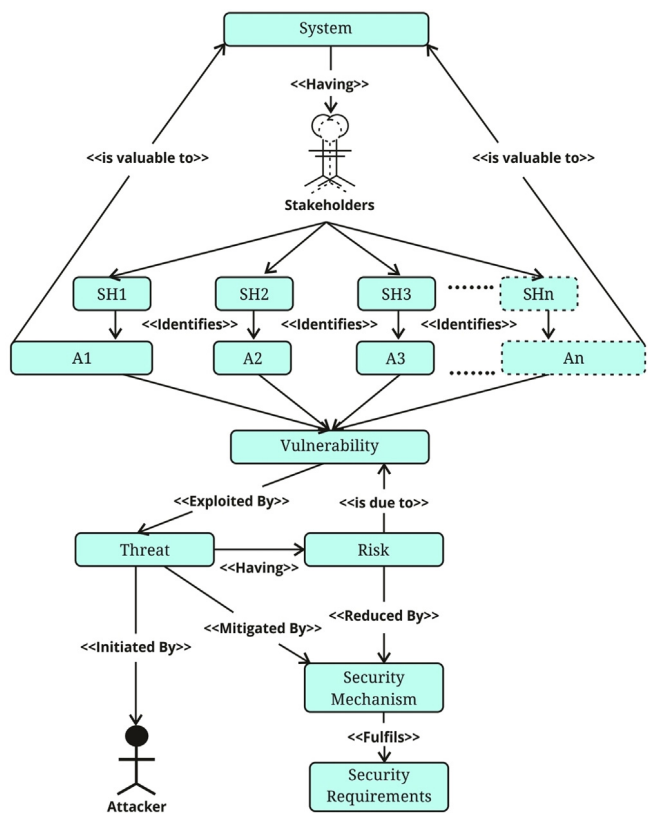

Fig. 1. Security requirement engineering concept model for STORE methodology.

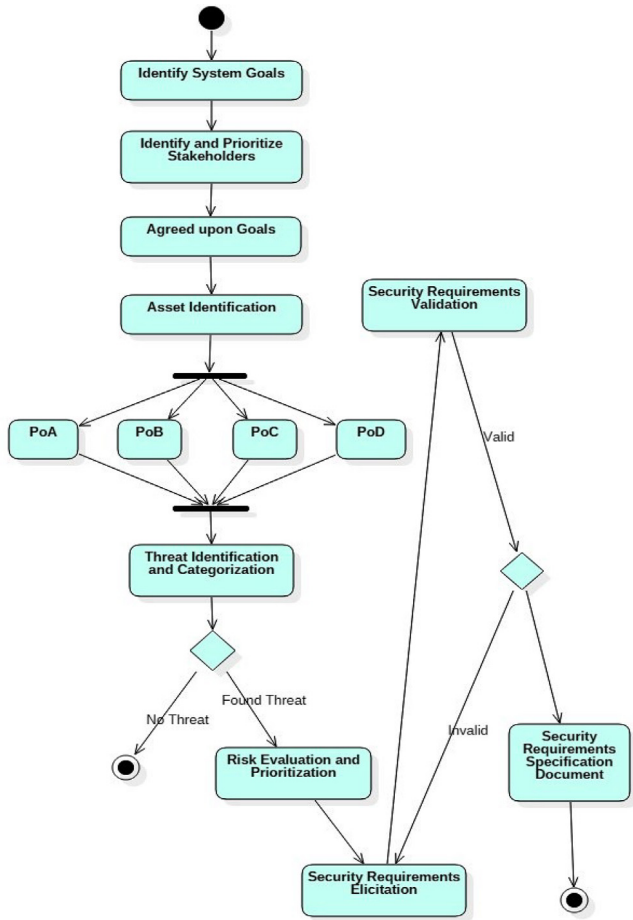

**Fig. 2.** Activity diagram of STORE Methodology.

### 3.2. Security Threat Oriented Requirements Engineering (STORE) methodology steps

The STORE Methodology is a ten-step security requirements engineering methodology. This section is to clarify the purpose of each step, the expectations of the participants, what will be the possible taking in and taking out for each step and also the appro-priate approach for each step. The following section lists all the steps of STORE methodology as well as describing the functionality of each step (Table 1).

    Step 1. Identify System Goals
    Step 2. Identify and Prioritize Stakeholders
    Step 3. Agreed upon Goals
    Step 4. Asset Identification
    Step 5. Security Attack Analysis

    A. Point of Attack (PoA)
    B. Point of Belief (PoB)
    C. Point of Conjecture (PoC)
    D. Point of Dependency (PoD)

    Step 6. Threat Identification and Categorization
    Step 7. Risk Evaluation and Prioritization
    Step 8. Security Requirements Elicitation
    Step 9. Security Requirements Validation
    Step 10. Security Requirements Specification Document

#### 3.2.1. Step 1. Identify system goals

Identify system goals is the first step of the proposed STORE methodology. Goals are written or verbal statements that illustrate the desired outcomes in a software product and the development team is aiming to accomplish it through several activities. The goals of a software system have to be identified by the requirement engineer from the client who wants software product. According to (Yue, 1987; Van Lamsweerde, 2001) goals are important compo-nents to identify in the requirements engineering process. Achiev-ing complete requirements is a major concern for any requirement engineer. Goals provide an accurate condition for getting satisfac-tory completion of a requirements specification. The outcome of requirement specification is complete with respect to a set of goals if all the goals can be proved to be achieved from the specification and the properties known about the domain considered. For better goal identification the preliminary analysis of the existing system is a vital starting place. The requirement engineer can use several techniques to identify system goals like interview, brainstorming etc. The output of the first step of STORE methodology is to produce a list of system goals.

**Table 1**
Steps in the proposed STORE Methodology.

| Step No. | Step | Taking in | Approach | Participants | Taking out |
|---|---|---|---|---|---|
| 1. | Identify System Goals | Objectives of the proposed system, Policy and Procedure | Interview, Brainstorming | Requirement Engineer, Client | List of System Goals |
| 2. | Identify and Prioritize Stakeholders | System Goals | Review, Analysis | Requirement Engineer | Stakeholders |
| 3. | Agreed upon Goals | System Goals | Meeting | Requirement Engineer, Stakeholders | Agreed upon Goals |
| 4. | Asset Identification | Stakeholder's Valuable Asset | Interview, questionnaire, brainstorming | Requirement Engineer, Stakeholders | Valuable Asset |
| 5. | Security Attack Analysis | Valuable Asset | Security Attack Analysis | Requirement Engineer, Security Expert | PoA, PoB, PoC, PoD |
| 6. | Threat Identification and Categorization | PoA, PoB, PoC, PoD | STRIDE Approach | Requirement Engineer | Categorized & Prioritized Threats |
| 7. | Risk Evaluation and Prioritization | Categorized & Prioritized Threats | DREAD Risk Assessment Methods | Requirement Engineer, Risk Manager | Risk Assessment Report |
| 8. | Security Requirements Elicitation | Potential Threats | Threat Dictionary | Requirement Engineer | Security requirements |
| 9. | Security Requirements Validation | Security requirements | Review, Walk-through | Requirement Engineer, Security Expert | Valid Security Requirements |
| 10. | Security Requirements Specification Document | Valid Security Requirement | Documentation | Requirement Engineer | Security Requirements Specification Document |

### 3.2.2. Step 2. Identify and prioritize stakeholders

The second step of STORE methodology is to identify and prioritize stakeholders who are associated with software system development. Several literature studies are available in the strategic management field which considers the importance of stakeholder by discussing organizations in provisions of a stakeholder model. (Sharp et al., 1999) described that identifying and prioritizing stakeholder is a vital process and this can be used to evaluate an organization's performance and control its future strategic direction. Several types of stakeholders are associated with any software development process. A stakeholder can express security concerns at different levels of detail (Fabian et al., 2010) According to (Glinz and Wieringa, 2007) "A stakeholder is a person or organization who influences a system's requirements or who are impacted by that system." A stakeholder can be a customer, end user, developer, requirement engineer, project manager and other persons who are associated with software development. All stakeholders are not equally important, therefore we must prioritize the identified stakeholders. The identified stakeholders may be prioritized into critical, major and minor. Each stakeholder has different security constraints to enforce the same service. Each stakeholder has their own security needs that may conflict with the other stakeholder's needs. (Almorsy et al., 2011) All stakeholders in a software development do not have equal importance, so (Glinz and Wieringa, 2007) prioritize the identified stakeholder roles with their importance.

- Critical: The stakeholder's role is critical if ignoring the stakeholder may destroy the project or render the system useless.
- Major: The stakeholder's role is major if ignoring the stakeholder would have a major unenthusiastic impact.
- Minor: The stakeholder's role is minor if ignoring the stakeholder would have a minor impact on the system.

### 3.2.3. Step 3. Agreed upon goals

Once the system goals and stakeholders based on their interests are identified, the next step of STORE methodology is to establish a proper understanding of system goals between the requirement engineers and different stakeholders. It is essential to articulate back this understanding to the stakeholders to get definition and agreement. The requirement engineers and stakeholders both must be agreed on the identified goals of the software product which is to be developed. This will create a perfect environment for the development of a software product in which every stakeholder can communicate with each other clearly. Definition and agreement with clear goals and objectives must be documented.

### 3.2.4. Step 4. Asset identification

Identifying valuable assets is one of the most important and complex steps of STORE methodology. The outcome of this step should be complete and correct otherwise this methodology will not elicit an effective and efficient security requirement. It is vital to identify all the valuable assets, in order to be able to elicit complete security requirements. Assets may be digital cash, data, password, information, commodities, people, computers etc. The assets should be identified with the help of all the potential stakeholders who are involved in the software development process. The requirement engineer can use different techniques like question-naires, brainstorming, interview, analysis, discussion for asset identification. The objective of security requirement engineering is to protect these valuable assets from the attacker so assets should be viewed not only in stakeholder perspective but also an attacker's point of view. Each stakeholder identifies assets from his point of view. Later the identified assets have to be categorized and prioritized. Further, the identified assets categorized under Confidentiality, Integrity, and Availability and prioritized as low, medium and high level of preference.

### 3.2.5. Step 5. Security attack analysis

Once the all valuable assets are identified and prioritized, the next step of STORE methodology is to analyze all the sources of security attack. Analyzing potential security attacks on a software system under development is an important step in engineering secure software systems, as the identified security attacks would elicit necessary security requirements. The following Fig. 3 shows the four points of security attack analysis that are essential for security requirements elicitation.

#### 3.2.5.1. A. Point of attack (PoA).
Point of attack for any software or web-based application is the point from where the adversary can enter into the system. The adversary always tries to identify PoA to attack the system. Point of attack represents like a loophole through which the adversary tries to interact with the system. A software has many points of attack like login page, sign up page, data entry page etc. Adversary tries to identify these points so he can potentially harm the system's confidentiality, integrity, and availability.

#### 3.2.5.2. B. Point of belief (PoB).
Point of belief shows the belief on external entities of the software system. The software system provides an access right to the external entities. These entities are known as the point of belief. Several entities can be the point of belief for a software product like the anonymous user who visit the web-based application, authorized users who have valid credentials, admin of software product, database server administrator etc. The security requirement engineer must identify all the point of belief while analyzing a security attack.

#### 3.2.5.3. C. Point of Conjecture (PoC).
Point of conjecture shows the hypothesis about security issues which may be faced by the software product in the future after the development. The Point of

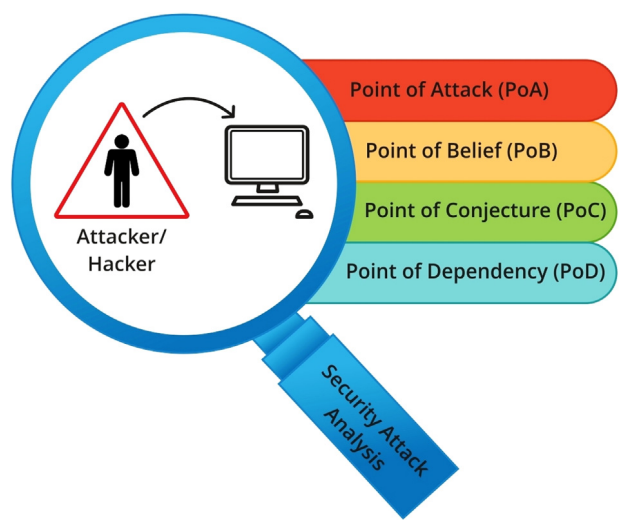

**Fig. 3.** Four points of security attack analysis.

Conjecture should be identified before the implementation phase has started and these points are very much significant in eliciting security requirements so this should not be despoiled. Point of Conjecture should be validated after the development of the software product.

*3.2.5.4. D. Point of Dependency (PoD).* Point of dependency shows the software dependency on external entities. For a software system, Point of Dependency could be the reason for potential harm. The PoD should be valid and authorized because inconsistency can show the way to a security attack. A software product can have several points of dependency like the Web server on which the web-based application depends, the database server, network connection etc.

### 3.2.6. Step 6. Threat identification and categorization

After identifying the four points PoA, PoB, PoC, PoD of security attack analysis in the previous step, it will be easy to identify the threats. This step identifies all potential threats to the software system which is to be developed. The threat is considered as a source of possible attacks to a software development and vulnerabilities are a weakness in the design of software. There are numerous potential threats that may harm the software application. Although it is not practical to identify all potential threats to a software development, all probable threats should be considered. There are a number of methods of identifying threats such as reviews of attack histories, reviews of current headlines involving data breaches, reviews of internet sites, Threat modeling. Threat modeling is an effective and best way to identify threats and vulnerabilities (Ashbaugh, 2008). Threat modeling is a technique used to look at the potential attacks that can be applied to a given software development by breaking down the software into its most basic components. We have constructed a threat dictionary that contains a collection of previously identified threats with corresponding most appropriate security requirements. The main intention is to help the security requirement engineer to elicit security requirements by accessing the threat dictionary knowledge. This step of STORE methodology also categorizes each threat through Microsoft's STRIDE model. STRIDE (Swiderski and Snyder, 2004) is a classification methodology for potential threats based on the threat categories like Spoofing, Tampering, Repudiation,

Information disclosure, Denial of service, and Elevation of privilege (Table 2).

### 3.2.7. Step 7. Risk evaluation and prioritization

The next step of STORE methodology is risk evaluation for identifying potential threats. This step evaluated the impact of threats on the software application. There are many risk assessment approach for assessing threats during the software development process. Mitigation of all the identified threats may not be necessary and economically feasible. The identified threats are ranked with their potential from the highest to lowest risk. The lowest rank threats may be ignored because of their low potential to harm the software system. A typical risk probability formula given below used in industry shows the risk and consequence of a particular vulnerability as equal to the probability of a threat occurring multiplied by the damage potential.

$$Risk = Probability \times Damage\ Potential$$

This formula measures the probability and damage potential on a 10 scale, where 1 represents the least likely to occur and 10 represents the most likely to occur. According to this formula if the probability of a threat occurring is 5 and damage potential of threat is 10 then the risk will be 50%. Risk evaluation can be done by Microsoft's DREAD risk assessment model (Swiderski and Snyder, 2004).

DREAD_Risk_Value = (DAMAGE + REPRODUCIBILITY + EXPLOITABILITY + AFFECTED USERS + DISCOVERABILITY)/5

Another risk assessment methodology is OCTAVE (Operationally Critical Threat, Asset, and Vulnerability Evaluation) is a very complex risk assessment approach developed by the SEI of CMU. Common Vulnerability Scoring System (CVSS) can also be used to rate the risk by the vulnerability. The Common Vulnerability Scoring System provides an open framework for communicating the characteristics and impacts of IT vulnerabilities (Mell et al., 2007). The Risk Evaluation and Prioritization step of STORE methodology calculate the impact of threats to assets and prioritize the threats with an inclusive evaluation about the degree of risk to an asset.

### 3.2.8. Step 8. Security requirements elicitation

The next step of STORE methodology is to elicit effective and efficient security requirements from the previous output of STORE methodology. The STORE methodology first considers the threats with higher risk, then the average risk potential threats and in last the low potential threats. We have proposed a threat dictionary that contains a collection of previously identified threats with corresponding most appropriate security requirements. The main intention is to help the security requirement engineer to elicit security requirements by accessing the threat dictionary knowledge. This step of STORE methodology elicited appropriate security requirements for each identified threat.

### 3.2.9. Step 9. Security requirements validation

After the successful elicitation of security requirements in the previous step of STORE methodology, we get efficient and effective security requirement for the mitigation of each potential threat. Threat mitigation is the security feature or assurance techniques for mitigating specific security threats. Unmitigated security threats are vulnerabilities that can be exploited by the attackers to harm the system. Every threat is not potentially harmful to the software system. Therefore, mitigation of every threat is not necessary.

### 3.2.10. Step 10. Security requirements specification document

The last step of STORE methodology is to create the SRS document. The security requirements specifications are made and they are validated with the stakeholders. This security requirement specification report will be submitted by the requirement engineer to the software development team for early integration of security in software development. (Finkelstein and Fuks, 1989) advocate that the construction of the specification document by taking several stakeholder viewpoints who have different views may be collected through a categorical dialogue that records responses in the form of statements, queries, rejections, challenges, etc.

**Table 2**
Microsoft's STRIDE Threat classification system.

| Acronym | Full form | Description |
|---|---|---|
| S | Spoofing | Spoofing threat allows an unauthorized user to become like an authorized user. In this category of threat, the attacker gets the log-in credentials of the authorized user through hacking techniques such as shoulder surfing, keystroke logging etc |
| T | Tampering | Tampering threat allows the attacker to modify the data within the software system to get the intentional malicious goal |
| R | Repudiation | This type of threat allows the attacker to claim that they didn't perform the malicious activity because the software system does not have sufficient evidence to prove otherwise |
| I | Information disclosure | In this type of threat, the attacker revealed the protected data, such as login credentials, credit/debit card number etc. to unauthorized users |
| D | Denial of service | This type of threat occurs when an attacker can prevent legitimate users from using the general functionality of the software system. |
| E | Elevation of privilege | This type of threat occurs when the attackers are able to gain additional rights and privileges for that they are not eligible |

## 4. Case study

### 4.1. Enterprise resource planning (ERP) system

The proposed STORE methodology applied to the Enterprise resource planning (ERP) web-based application software. This college ERP system is capable of managing student records, department, faculties, library, and other information. The college ERP system has all the information about the students, faculties, staff, library, departments and other confidential information which requires security. The proposed methodology was applied to this college ERP system before the development of this software product so that the requirement engineer can easily elicit all the security requirements needed for this software product. Eliciting the entire security requirement before the development of a software product is not an easy task, but the proposed methodology makes it easy for the requirement engineer. Several literature studies have suggested several significant success factors in ERP implementation (Nah and Lau, 2001; Hong and Kim, 2002; Umble et al., 2003; Al-Mashari et al., 2003; Al-Mashari and Al-Mudimigh, 2003; Okunoye et al., 2008). For the successful implementation of an ERP system requirement must be created and the system requirement should be documented in a proper manner (Scheer and Habermann 2000). According to Nah and Lau (2001), an ERP system has the ability to automate and integrate an organization's business processes, share common data and practices across the entire enterprise and produce and access information in a real-time environment. An ERP system can be highly beneficial to an organization if the organization is able to overcome the implementation hurdles. ERP systems cannot remain inside organizational boundaries. Technically, ERP developers are preferably considering the browser/Web server architecture over traditional client/server in order to deliver e-business capabilities. Hence, for their first-generation e-business solutions almost all big ERP vendors are using a mixed Java/XML strategy (Scheer and Habermann, 2000).

### 4.2. Execution of the case study

A case study of an ERP system presented here shows the support of the STORE methodology for security requirements engineering. Execution of the case study for the validation of a new methodology is a very common and complex task in the software engineering field. All the sequential steps of STORE methodology have been followed in the following section for the effective and efficient security requirements engineering.

### 4.2.1. Step 1. Identify system goals

Identify system goals is the first step of the proposed STORE methodology. In this step, the requirement engineer collects all the system goals for the software system which is to be developed. Requirement engineer can use several techniques to identify and collect system goals like interviews, brainstorming, etc. For the college ERP system following can be system goals (see Table 3).

**Table 3**
Identified system goals for college ERP system.

| Goal ID | Description |
|---|---|
| G1 | The college ERP system will be installed on a web server that has been secured to protect confidential information. All security patches for the web server must be enabled |
| G2 | The college ERP system will also be installed on a database server that has been secured. All security patches for the database server must be enabled |
| G3 | The database server must be protected from the direct access from the internet by a firewall |
| G4 | The Web server should be protected from direct access from the internet by a firewall |
| G5 | Only HTTP and HTTPS ports allowed direct access from the internet |
| G6 | Communication between the web server and database server should be conducted over a private network |
| G7 | The college ERP system should be deployed over HTTPS |

### 4.2.2. Step 2. Identify and prioritize stakeholders

This step of STORE methodology, identify and prioritize the internal and external stakeholders concerned at different levels of ERP system development (Table 4).

### 4.2.3. Step 3. Agreed upon goals

After identifying all the stakeholders related to the ERP system project the next step of STORE methodology is to make all the stakeholders agreed upon the proposed system goals. The requirement engineers and stakeholder both must be agreed on the identified goals of the software product for the development. This will create a perfect environment for the development of a software product in which every stakeholder can communicate with each other clearly.

### 4.2.4. Step 4. Asset identification

The STORE methodology is used to protect the assets from threats. This step is an important step of the STORE methodology as this step identifies the assets related to a particular software system (Table 5).

### 4.2.5. Step 5. Security attack analysis

After identifying all the valuable assets, the next step of STORE methodology is to analyze all the sources of security attacks to the college ERP system. Analyzing potential security attacks on ERP software system under development is an important step in engineering secure software systems, as the identified security attacks will help in identifying important security threats. The following four points are essential for security attack analysis.

#### 4.2.5.1. A. Point of attack (PoA).
The following table shows the identified Point of Attack (PoA) for the college ERP system in a security attack analysis process of STORE methodology (Table 6).

**Table 4**
List of possible stakeholders with their significance for college ERP system.

| No. | Name | Significance | Type |
|---|---|---|---|
| 1 | President | Critical | Managerial |
| 2 | Director | Critical | |
| 3 | Senior Executives | Major | |
| 4 | Internal Auditor | Critical | Marketing |
| 5 | Purchasing Manager | Critical | |
| 6 | Key users | Major | |
| 7 | End users | Critical | |
| 8 | ERP Project manager | Critical | Information System |
| 9 | Database administrator | Critical | |
| 10 | Developer | Critical | |
| 10 | Networking team | Major | |

**Table 5**
Identified assets of the college ERP system.

| Asset ID | Name | Description |
|---|---|---|
| A1 | Student, staff, and admin | An asset that relates to a student, staff or admin |
| A2 | Student's login data | The student's credentials: username and password |
| A3 | Staff login data | The staff's credentials: username and password |
| A4 | Admin login data | The admin's credentials: username and password |
| A5 | Student's personal data | The personal data that the student enters, such as student record and assets |
| A6 | Staff's personal data | The personal data that the staff enters, such as staff record and assets |
| A7 | System | Assets that relate to the essential system |
| A8 | Availability of ERP System | If the college ERP system goes down, student/and staff cannot request or receive quotes |
| A9 | Process | Assets that relate to the process of running the web application |
| A10 | Application | Assets that relate to the web application |
| A11 | Login Session | The web session associated with a logged in student, staff or admin |
| A12 | Backend database access | The ability to interact with the database that stores, student's data, staff data, and login credentials |
| A13 | Student fee details | The student's fee record must be secure. Tampering with this data could cause the loss of college |
| A14 | Staff salary details | The staff salary record must be secure. Tampering with this data could cause the loss of college |
| A15 | Message Notification | The message notification contains the information for students and staff |
| A16 | Audit data | Attackers might try to attack the system without being logged or audited |
| A17 | Access to the record | Only authorized people should be able to view his/her record |

**Table 6**
List of identified Point of Attack (PoA).

| ID | Point of Attack | The port (HTTPS) that the web server listens on |
|---|---|---|
| PA1 | Web Server Listening Port (HTTPS) | The port (HTTPS) that the web server listens on |
| PA2 | Login Page | Page for students or staff to create a login and perform a login to the site to begin requesting or reviewing records |
| PA3 | CreateLogin function | Creates a new student or staff login (Admin login must be created directly through the database stored procedures.) |
| PA4 | LoginToSite function | Compares authorized person credentials to those in the database and if credentials match, create a new session |
| PA5 | Data entry page | Page used to enter student or staff personal data into the database so that the admin can review it |
| PA6 | RetrieveData function | Allow the authorized person to view his/her records from the database |
| PA7 | SubmitData function | Submits student or staff data to be reviewed by the admin |
| PA8 | Admin Review page | This page used by the admin to review the student or staff request |
| PA9 | RetrieveData function | Retrieves student or staff data |
| PA10 | SubmitData function | Submits any information for the student or staff |
| PA11 | ListRequests function | Lists requests ready for review. |
| PA12 | Database Listening Port | Enables the database to be used remotely by the authorized persons |
| PA13 | Database stored procedures | Store and retrieve records in the database |
| PA14 | CreateLogin procedure | Create a login for the authorized person |
| PA15 | RemoveLogin procedure | Logout from the college ERP system |
| PA16 | StoreUserData procedure | Used to store user data from the data entry page of the ERP system |
| PA17 | RetrieveUserData procedure | Retrieves the user's data and request |

*4.2.5.2. B. Point of belief (PoB).* The following table shows the identified Point of Belief (PoB) for the college ERP system in a security attack analysis process of STORE methodology (Table 7).

*4.2.5.3. C. Point of Conjecture (PoC).* The following table shows the identified Point of Conjecture (PoC) for college ERP system in a security attack analysis process of STORE methodology (Table 8).

*4.2.5.4. D. Point of Dependency (PoD).* The following table shows the identified Point of Dependency (PoD) for the college ERP system in a security attack analysis process of STORE methodology (Table 9).

#### 4.2.6. Step 6. Threat identification

After identifying the assets and analyzing the security attack with the help of STORE methodology, it is necessary to identify the potential threat for each asset. The following table consists of recognized threats to a college ERP system (Table 10).

#### 4.2.7. Step 7. Risk Evaluation and prioritization

The Risk assessment and prioritization step of STORE methodology calculate the impact of threats to assets, and shows an inclusive evaluation of the degree of risk to an asset. The following table consists of the identified threats with their risk value (Table 11).

#### 4.2.8. Step 8. Security requirements elicitation

This step of STORE methodology elicits the appropriate security requirement for each identified potential threats. The following table consists of security requirements for each identified threat of the college ERP system (Table 12).

**Table 7**
List of identified Point of Belief (PoB).

| ID | Point of Belief | Description |
|---|---|---|
| PB1 | Unauthorized remote user | A user who has connected to the ERP system, but has not provided valid credentials yet |
| PB2 | Authorized remote user | An authorized user who has created an account and has valid login credentials |
| PB3 | Admin | Admin uses login credentials to access and modify the database |
| PB4 | HTTP user | A remote user that accesses a page via HTTP |
| PB5 | HTTPS user | A remote user that accesses a page via HTTPS |
| PB6 | Web server process identity | Used to authenticate the web server to the database when storing or retrieving information |
| PB7 | Database server process identity | The account that the database server process runs as, represented by its process token |

**Table 8**
List of identified Point of Conjecture (PoC).

| ID | Point of Conjecture | Description |
|---|---|---|
| PC1 | Online payment | The online payment system can be another function for this ERP system. If this functionality added, this function should not provide a way for attackers to attack existing security features |
| PC2 | Payment Gateway | If added payment gateway in future, ERP system must comply with PCI DSS or other security standards |
| PC3 | Encrypted Communication | If encrypted communication functionality is added to the ERP system in the future, message exchange should be completed according to standards |

**Table 9**
List of identified Point of Dependency (PoD).

| ID | Point of Dependency | Description |
|----|---------------------|-------------|
| PD1 | Database Server | The ERP system depends on the security of the database server |
| PD2 | Web Server | The ERP system depends on the security of the web server |
| PD3 | Network | The ERP system depends on the security of the network between the web server and database server |
| PD4 | External SMTP | The ERP system depends on an external SMTP server to deliver any message |
| PD5 | Session Management | The ERP system depends on the session management of the web server being secure |

#### 4.2.9. Step 9. Security requirement validation

This step is related to the validation of recognized security requirements for college ERP system obtained in the previous step of the STORE methodology. All the identified security requirements must be capable to enhance the quality of the software system by enforcing appropriate security concern. In this step, the security requirements are prioritized based on the potential risk of the associated threat. The validation of the security requirements process consists of security requirement engineers and concern stakeholders as a participant. They use the review or walk-through approaches to validate the security requirements for the college ERP system.

#### 4.2.10. Step 10. Security requirements specification document

Generating security requirement specification document is the last and crucial step of STORE methodology. All recognized and validated security requirement is documented in an organized manner for the college ERP system which is considered as a web-based application software. This security requirements specification document assists the developer to develop a more secure and quality ERP system which ensures security from the beginning of the software product.

### 5. Result and discussion

This section describes the comparative analysis of our proposed STORE methodology with two different security requirements elicitation approaches that are similar to our proposed methodology.

We have compared STORE methodology with SQUARE (Gordon et al., 2005) and MOSRE framework (Salini and Kanmani, 2012). We have chosen to compare with these security requirements engineering techniques because our proposed STORE methodology is motivated by such methodologies in the way of eliciting security requirements for the software products. We showed that the security requirements are more effective and efficient when following the sequential steps of our proposed STORE methodology.

#### 5.1. Comparison with the SQUARE methodology

In this section, we have applied our proposed STORE methodology in a case study presented by Gordon et al. (2005). They have applied the SQUARE methodology for eliciting security requirements on a case study of the Asset management system. They elicited total nine security requirements and after requirements prioritization, they selected five security requirements R01, R02, R06, R07, R08 out of nine as essential security requirements. The following table shows the comparison between SQUARE methodology and STORE methodology. We have compared our results with the essential security requirements obtained by SQUARE methodology. The result shows that the STORE methodology covers more threats than SQUARE methodology and also elicits more effective and efficient security requirements. Therefore, STORE methodology elicits more complete security requirements for the asset management system (Table 13).

#### 5.2. Comparison with the MOSRE framework

We have also applied our proposed STORE methodology in the case study presented by the Salini and Kanmani (2012). They have applied their proposed MOSRE framework for eliciting security requirements on a case study of the E-Health system. They applied MOSRE in the early stages of E-Health system development, to identify assets, threats, and vulnerabilities. The following table shows the comparison between MOSRE framework and STORE methodology. We have compared our results with the system asset-based security requirements obtained by MOSRE framework. The result shows that the STORE methodology covers more threats than MOSRE framework and also elicits more effective and efficient security requirements for the E-Health system. Therefore, STORE methodology elicits more complete security requirements (Table 14).

**Table 10**
Identified threats for the college ERP system.

| ID | Threat | Description | STRIDE | | | | | | Mitigated | Assets |
|----|--------|-------------|---|---|---|---|---|---|-----------|--------|
| | | | S | T | R | I | D | E | | |
| T1 | Malicious SQL data in user input | The attacker might try to inject SQL commands into the application via Login. | | ✔ | | | | ✔ | No | A12 |
| T2 | Login Information Disclosure | The attacker gets the login credentials of the authorized user. | | | | ✔ | | ✔ | No | A2, A3, A4 |
| T3 | Session Id Theft | The attacker gets the session ID of another authorized user. | | | | | | ✔ | No | A11 |
| T4 | User Data Disclosure | Disclosing another authorized user data raises privacy issues. | ✔ | | | ✔ | | | No | A5, A6 |
| T5 | Access to the Database | The Attacker attacks to the database of the ERP system. | | ✔ | ✔ | ✔ | | | Yes | A1-A6 |
| T6 | Attack on Admin Login | The attacker performs as an admin of the ERP system. | | | | | | ✔ | Yes | A4 |
| T7 | Blocking Message Notification | The attacker prevents an authorized user from receiving any notification. | | | | | ✔ | | Yes | A15 |
| T8 | User Data Tampering | The attacker modifies the authorized person's data. | ✔ | ✔ | | | | ✔ | No | A5, A6 |
| T9 | User Account Deletion | The attacker deletes an authorized user account. | | | | | ✔ | ✔ | Yes | A2, A3, A4 |
| T10 | Crashing the ERP system | The attacker crashes the ERP web application. | | | | | ✔ | | Yes | A8 |
| T11 | Unauthorized access | The attacker access the ERP system without valid credentials. | | | | | | ✔ | Yes | A5, A6 |
| T12 | Access without Login | The attacker access the information of authorized person without being logged. | | | | ✔ | | | No | A16 |

**Table 11**
List of prioritize threats to their DREAD risk value.

| Threat ID | Threat | DREAD Value | Mitigated |
|---|---|---|---|
| T1 | Malicious SQL data in user input | 10 | No |
| T5 | Access to the Database | 10 | Yes |
| T10 | Crashing the ERP system | 10 | No |
| T4 | User Data Disclosure | 9.2 | No |
| T8 | User Data Tampering | 9.2 | No |
| T6 | Attack on Admin Login | 7.6 | Yes |
| T9 | User Account Deletion | 7.6 | Yes |
| T12 | Access without Login | 7.6 | No |
| T2 | Login Information Disclosure | 6.6 | No |
| T7 | Blocking Notification | 6.4 | Yes |
| T11 | Unauthorized access | 5.2 | Yes |
| T3 | Session Id Theft | 3.8 | No |

**Table 12**
List of elicited security requirements for every threat.

| Threat ID | Security Requirement ID | Security Requirement |
|---|---|---|
| T1 | SR1 | Use of prepared statements with parameterized queries |
| T5 | SR2 | Use of Access control, Auditing, Authentication, Encryption, Integrity controls, Backups techniques |
| T10 | SR3 | Upgrade to the new version by fixing all identified flaws |
| T4 | SR4 | Use of complex encryption methods that limits the risks of user data disclosure of ERP system |
| T8 | SR5 | Use a firewall and proper authorization technique for granting the access right to use of the software system |
| T6 | SR6 | Implement account lockout procedure, captcha and enforce the user of the ERP system to use strong passwords |
| T9 | SR7 | Complex security password and account lockout should be used which locked the account after some failed login attempts |
| T12 | SR8 | Use firewalls, VPN and SSL techniques |
| T2 | SR9 | The database server of ERP system should be protected from the direct internet access by a firewall |
| T7 | SR10 | Ensure the proper security of SMTP server |
| T11 | SR11 | Implement two-factor authentication, i.e. strong password and one-time passcode |
| T3 | SR12 | Use SSL/HTTPS encryption for the ERP system |

**Table 13**
The comparison results with the SQUARE methodology.

| *SQUARE Methodology results* | |
|---|---|
| Essential security requirements | |
| R01 | The system is required to have strong authentication measures in place at all system gateways/entrance points |
| R02 | The system is required to have sufficient process-centric and logical means to govern which system elements (data, functionality, etc.) users can view, modify and/or interact with |
| R06 | It is required that the system's network communications be protected from unauthorized information gathering and/or eavesdropping by encryption and other reasonable techniques |
| R07 | It is a requirement that both process-centric and logical means be in place to prevent the installation of any software or device without prior authorization |
| R08 | It is required that the AMS's physical devices be protected against destruction, damage, theft, tampering or surreptitious replacement (including but not limited to damage due to vandalism, sabotage, terrorism or acts of God/Nature) |
| *Results from the proposed STORE methodology* | |
| Elicited Security Requirements | |
| SR1 | Use of Access control, Auditing, Authentication, Encryption, Integrity controls, Backups techniques |
| SR2 | Implement account lockout procedure, captcha and enforce the user of the ERP system to use strong passwords |
| SR3 | Use of complex encryption methods that limits the risks of user data disclosure of E-Health system |
| SR4 | Use a firewall and proper authorization technique for granting the access right to use of the software system |
| SR5 | Use HIPAA security standards and policy to ensure proper external security |

**Table 14**
The comparison results with MOSRE framework.

| *MOSRE framework results* | |
|---|---|
| System asset-based security requirements | |
| SR1 | Use secure authentication, which does not send passwords over the network |
| SR2 | Use secure communication channels |
| SR3 | Use remote procedure call encryption |
| SR4 | Firewall policies that block all traffic except expected communication ports |
| *Results from the proposed STORE methodology* | |
| Elicited Security Requirements | |
| SR1 | Use of Access control, Auditing, Authentication, Encryption, Integrity controls, Backups techniques |
| SR2 | Implement account lockout procedure, captcha and enforce the user of the E-Health system to use strong passwords |
| SR3 | Use of complex encryption methods that limits the risks of user data disclosure of E-Health system |
| SR4 | Use a firewall and proper authorization technique for granting the access right to use of the software system |

## 6. Conclusion

In this paper, we have described our proposed approach, Security Threat Oriented Requirement Engineering (STORE) methodology for effective and efficient elicitation of security requirements. The STORE methodology overcomes the several limitations of other existing approaches like they don't categorize or prioritize the potential threats, lack of coding standards, process planning and unorganized documentation of security requirements etc. of several other security requirements elicitation approaches. These issues may lead to complexity of adopting the other framework. The existing methodology follows the iterative process which may add advantage to find new security requirements but this will lead to more complexity to the web applications. The simplicity of the STORE methodology makes it usable by technical persons and developers who are not expert in the software security field. STORE methodology integrates security with the requirements engineering process based on threats. STORE methodology identifies and priorities the potential threats and eliciting the most appropriate security requirements in a systematic way.

The identification and analysis of four points of security attack PoA, PoB, PoC, PoD makes the identification of threats easier. We have constructed a threat dictionary that contains a collection of previously identified threats with corresponding most appropriate security requirements. The main intention is to help the security requirement engineer to elicit security requirements by accessing the threat dictionary knowledge. We have also successfully validated our proposed STORE methodology by applying this approach for a case study of ERP system. Hence, we have designed and developed a security requirement elicitation methodology that is easy to adopt, extra comprehensive and assist the requirement engineers to elicit effective and efficient security requirements in a more organized manner. STORE methodology can be used for the web applications which consider information as treasure or assets and evaluated the strength of the methodology. In the early phases of STORE methodology, we begin the context of the system, the functional requirements, and the primary security goals and requirements. The STORE methodology approach presented here has not yet been validated on a big project. The next step is to deploy it on a real time project. We want to ensure that our approach is practical, and a suitably accurate assessment requires a real project. In the future, we have planned to apply our STORE methodology for many other software development projects. The other future work is to develop a tool for STORE methodology to elicit security requirements.